# Electronic inhomogeneities in graphene: the role of the substrate interaction and chemical doping

## Inhomogeneidades electrónicas en grafeno: el rol de la interacción con el substrato y el dopaje químico

A. Castellanos-Gomez[1,2,+,*], Arramel[2], M. Wojtaszek[2], R.H.M. Smit[1,3], N. Tombros[2], N. Agraït[1], B.J. van Wees[2], G Rubio-Bollinger[1]

[1] Física de la Materia Condensada. Universidad Autónoma de Madrid, Campus de Cantoblanco, E-28049 Madrid, Spain.
[2] Physics of Nanodevices, Zernike Institute for Advanced Materials, University of Groningen, The Netherlands.
[3] Kamerlingh Onnes Laboratorium, Leiden University, P.O. Box 9504, 2300 RA Leiden, The Netherlands.
[+] Present address: Kavli Institute of Nanoscience, Delft University of Technology, Lorentzweg 1, 2628 CJ Delft, The Netherlands.
[*] Corresponding author: a.castellanosgomez@tudelft.nl

**Abstract**

We probe the local inhomogeneities of the electronic properties of graphene at the nanoscale using scanning probe microscopy techniques. First, we focus on the study of the electronic inhomogeneities caused by the graphene-substrate interaction in graphene samples exfoliated on silicon oxide. We find that charged impurities, present in the graphene-substrate interface, perturb the carrier density significantly and alter the electronic properties of graphene. This finding helps to understand the observed device-to-device variation typically observed in graphene-based electronic devices. Second, we probe the effect of chemical modification in the electronic properties of graphene, grown by chemical vapour deposition on nickel. We find that both the chemisorption of hydrogen and the physisorption of porphyrin molecules strongly depress the conductance at low bias indicating the opening of a bandgap in graphene, paving the way towards the chemical engineering of the electronic properties of graphene.

**Resumen**

Hemos estudiado las inhomogeneidades locales de las propiedades electrónicas del grafeno a escala nanométrica utilizando técnicas de microscopía de sonda próxima. En primer lugar, nos centramos en el estudio de las inhomogeneidades electrónicas causadas por la interacción del grafeno con el sustrato en muestras de grafeno exfoliado sobre óxido de silicio. Encontramos que las impurezas cargadas, presentes en la interfaz entre el grafeno y el sustrato, perturban considerablemente la densidad de portadores y alteran las propiedades electrónicas del grafeno. Este hallazgo ayuda a comprender la gran variabilidad entre distintos dispositivos que se observa típicamente en dispositivos electrónicos basados en grafeno. En segundo lugar, investigamos el efecto de la modificación química de las propiedades electrónicas de grafeno, crecido sobre níquel por depósito por vapor químico. Encontramos que tanto la quimisorción de hidrógeno como la fisisorción de moléculas de porfirina logran reducir fuertemente la conductancia a bajo voltaje, lo que indica la apertura de un *gap* en el grafeno, allanando el camino hacia el diseño químico de las propiedades electrónicas de grafeno.

## 1. Introduction

Since the first experimental realization of graphene [1], its unique properties [2-4] have boosted the research in this novel material. Due to the high sensitivity of graphene to external electric fields, it can be used for sensing applications or as the channel in field effect transistors. This high sensitivity, however, makes graphene very vulnerable to charged impurities in the surroundings and to chemical doping [5, 6], leading to a large device-to-device variation in the electronic performance [6]. This effect is especially critical because graphene is all surface and thus when it lays on a surface or it is covered by a layer of adsorbates almost every single carbon atom of the graphene layer can be altered by the graphene-substrate interaction or the chemical doping.

In this work we review experimental results on the local electronic properties of graphene, focused first on the study of the electronic inhomogeneities caused by the graphene-substrate interaction (based on Ref. [7]) and second on the variation of the local electronic properties of graphene by chemical modification of the surface (based on Ref. [8]).

## 2. Graphene-substrate interaction: exfoliated graphene on $SiO_2$

When graphene is exfoliated on insulating substrates, the presence of charged impurities typically generates electric fields strong enough to change the doping level of the graphene layer at the nanoscale [9, 10]. As a consequence, graphene-based electronic devices show a large device-to-device variation in electrical performance and reproducibility [5, 6].
We have developed a combined scanning tunnelling and atomic force microscope (STM/AFM) to characterize the electronic properties of graphene layers even when they are deposited on top of insulating substrates. This scanning probe microscopy (SPM) tool can operate as an AFM, without the need of a conductive substrate, to locate the graphene flake. The microscope relies on an STM which has been supplemented with a piezoelectric quartz tuning fork force sensor [11, 12] in the so-called *qPlus* configuration [13] (spring constant $k \sim 12500$ N/m, resonance frequency $f_0 \sim 32.1$ kHz and quality factor $Q \sim 4200$). The use of a carbon fiber tip has found to optimize the performance of this type of combined STM/AFM microscopes [14-16].

The measurement starts by positioning the carbon fiber tip on top of a graphene flake (prepared by cleavage of highly oriented pyrolytic graphite on a $SiO_2$/Si substrate using silicone stamps [17]) with the help of a long working distance optical microscope (see Figure 1a). Then the AFM capability of our combined STM/AFM is used to scan the region under study (see Figure 1b) and thus to determine thickness of the different areas of the flake. Before starting the STM measurements the tip is positioned onto the flake, which is electrically contacted by a gold electrode (deposited by shadow mask evaporation), and the scan range is reduced (dashed square in Figure 1b) in order to avoid the tip reaching the insulator substrate which would result in a tip crash that degrades the sample and/or tip irreversibly. Figure 1c shows the topography of a mechanically exfoliated graphene flake deposited on a 285 nm $SiO_2$/Si substrate obtained in the constant current STM mode. All STM measurements were acquired under ambient conditions.

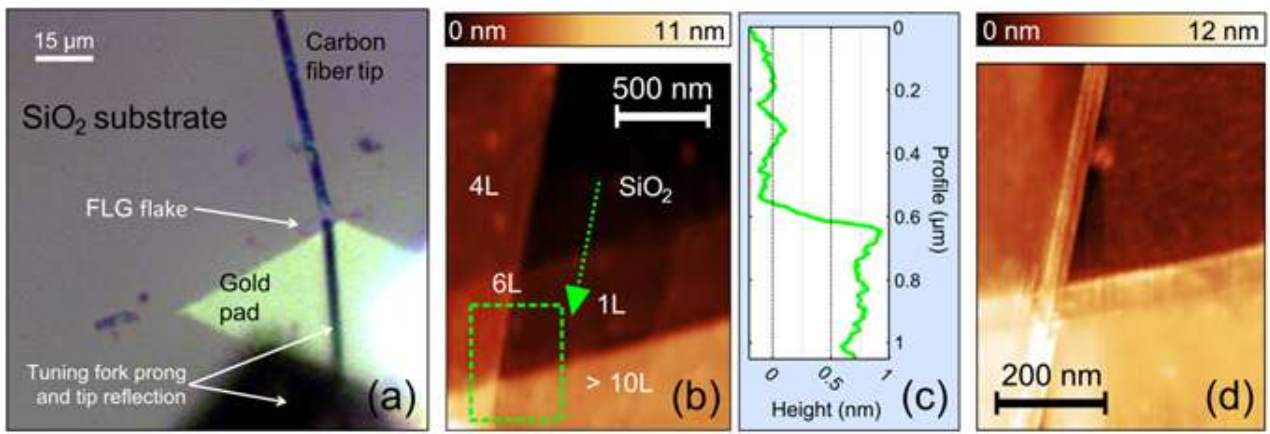

**Figure 1.** (a) Optical image of the coarse tip positioning on a few-layers graphene flake. (b) AFM topography image of the interface between the few-layers graphene flake and the $SiO_2$ substrate. Areas with different number of layers (labelled as >10L, 6L, 4L and 1L) are found. (c) Topographic line profile acquired along the dotted line in (b), showing the interface between the $SiO_2$ substrate and a monolayer graphene region. (d) STM topography image of the region marked by the dashed rectangle in (b). Adapted from Ref. [7].
**Figura 1.** (a) Imagen óptica del posicionamiento grueso de la punta del microscopio sobre un copo de pocas capas de grafeno. (b) Imagen de topografía AFM imagen de la interfaz entre el copo de pocas capas de grafeno y el sustrato $SiO_2$. Se distinguen áreas con diferente número de capas (etiquetadas como > 10L, 6L, 4L y 1L). (c) Perfil topográfico adquirido a lo largo de la línea de puntos en (b), mostrando la interfaz entre el sustrato $SiO_2$ y una región monocapa de grafeno. (d) Imagen topográfica STM de la región marcada por el rectángulo de trazos en (b). Adaptada de Ref. [7].

In order to gain a deeper insight in the role of charged impurities on the electronic properties of graphene, we have also measured (simultaneously with the topography) the variations of the tunneling barrier height (see the supplementary information of Ref. [7] for technical details) because the tunnelling current decay constant ($\beta$) is strongly influenced by the presence of subsurface charges [18, 19]. In fact, the presence of negatively (positively) charged impurities causes an electric field which effectively shifts upwards (downwards) the energy of the bottom of the band by an amount $\Delta E$. This effect reduces (increases) the apparent tunneling barrier height, $\Phi_{app}$. This barrier height change can be modeled as $\Phi_{app} = (\Phi_{graphene} - \Phi_{tip} \pm \Delta E)$ where $\Phi_{graphene}$ and $\Phi_{tip}$ are the graphene and the tip work function respectively. The associated tunnelling decay constant then changes according to $\beta = 2\sqrt{2m \cdot \Phi_{app}} / \hbar$ ($m$ is the electron mass and $\hbar$ is the reduced Planck constant).

Figure 2a shows the spatial variation of $\beta$ for a multilayer graphene simultaneously measured with the STM topography (Figure 1d). The $\beta$ image for a multilayer shows an almost constant value (5.3 ± 0.5 nm$^{-1}$), typical of STM operation in air [20]. The spatial variation of $\beta$ for single-layer graphene, on the other hand, shows strongly localized inhomogeneities (identified as dips in the $\beta$ image) caused by local doping induced by negative charged impurities in the substrate (Figure 2b). These inhomogeneities in the tunnelling decay constant can be also observed by measuring tunnelling vs. distance traces at different positions in the single layer graphene region (Figure 2d).

Considering that each dip in the $\beta$ image (Figure 2b) is due to the presence of one individual negative charge we can estimate the density of charged impurities by tentatively counting the number of depressions in a given area, resulting in an impurity density $\sigma = (2.9 \pm 0.6) \cdot 10^{11}$ cm$^{-2}$. A more objective procedure to statistically analyse the short-range ordering [21] and spatial variation of $\beta$ is to obtain the radially-averaged autocorrelation function $g(r)$, shown in Figure 5c. The typical radius $r_{FWHM}$ at half minimum of the localized inhomogeneities is obtained from the radial distance at which the value of $g(r)$ is 0.5, which yields $r_{FWHM}$ = 4.7 ± 1.1 nm. The inhomogeneities do not show long range ordering and the mean spacing between them is $d$ = 22 ± 2 nm, determined from the position of the first maximum of $g(r)$[21], which corresponds to a charge density $\sigma = \pi^{-1}(d/2)^{-2} = (2.6 \pm 0.5) \cdot 10^{11}$ cm$^{-2}$. As the radius of the ß dips presents a well-defined average value, it indicates that the charged inhomogeneities are localized at a well-defined distance from the graphene layer. If the charged impurities would be located randomly in the $SiO_2$ layer (300 nm thick) the distribution of radius would be much broader. Therefore, it is reasonable to assume that the charged impurities are located at the interface between the $SiO_2$ layer and the graphene flake.

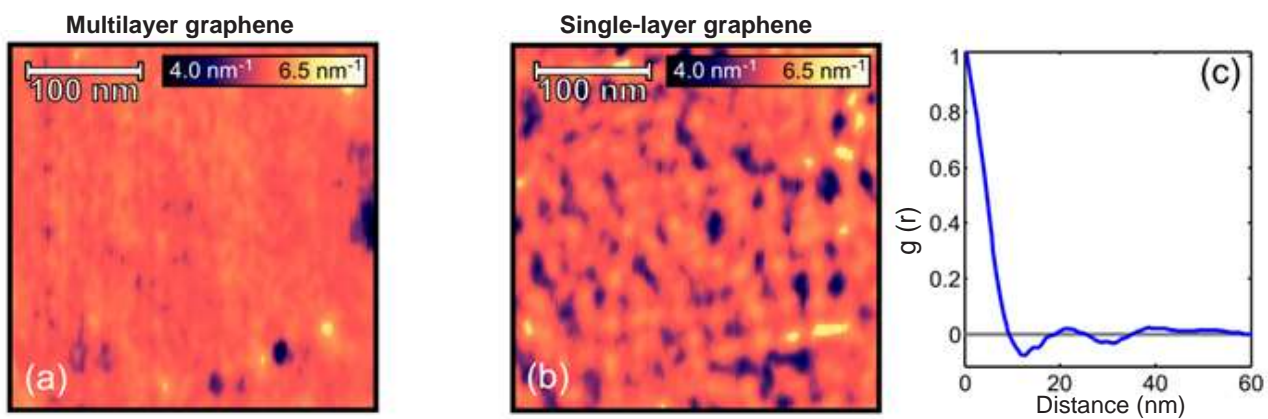

**Figure 2.** (a) and (b) show the local tunnelling decay constant maps measured on a multilayer and a single-layer region, respectively. (c) Radial autocorrelation function of the local tunnelling decay image in (b). Adapted from Ref. [7].
**Figura 2.** (a) y (b) muestran los mapas de la constante de decaimiento túnel medidos en una región de multicapa y una región de una sola capa, respectivamente (c) Función de autocorrelación radial de la imagen de la constante de decaimiento túnel mostrada en (b). Adaptada de Ref. [7].

## 3. Effect of chemical doping on the local electronic properties of CVD graphene

Chemical functionalization of graphene is a very promising route to achieve large scale production of semiconducting graphene for industrial applications. Indeed, the application of graphene in microelectronic devices is hampered by the lack of a bandgap which is mandatory in order to fabricate electrically switchable devices with large current on-off ratio.

Up to date several strategies have been developed to open a band gap in graphene. Here we have explored two different options. First, the exposure of the graphene surface to hydrogen plasma [22-24] that leads to the chemisorption of hydrogen atoms, which produces the sp$^3$ hybridization of carbon network, reduces the number of delocalized sp$^2$ electrons and consequently opens a band gap. Second, the non-covalent stacking of aromatic organic molecules on graphene through π-π interaction.

### 3.1. Hydrogen chemisorption

The hydrogenation of CVD graphene has been done using Ar:$H_2$ plasma (composition of 85:15) in a reactive ion etching system with a high frequency generator operating at 13.56 MHz, capacitively coupled to the bottom electrode [25]. The gas flow is kept constant at 200 sccm and the pressure in the chamber is 0.05 mbar. We chose the lowest plasma ignition power, P = 3 W (power density is ~ 4 mW/cm$^2$), and tuned the circuit impedance to reduce the built-in DC self-bias between the bottom electrode and the plasma down to zero. Each exposure is done for 40 min, which leads to moderate hydrogen coverage and is directly followed by STM/STS measurements.

Figure 3 shows an 9 × 9 nm$^2$ topographic image acquired in the constant current STM mode of the surface of graphene grown on nickel by CVD.[26] Before the plasma treatment, the CVD graphene exhibits a Moiré pattern superimposed to the honeycomb lattice of graphene (Figure 3a). This is due to the lattice parameter mismatch between the graphene and the nickel surfaces and it is the characteristic of most of epitaxial graphene samples [27-29].

On the other hand, for the hydrogenated CVD graphene, the expected structural changes are twofold. First, the chemisorption of hydrogen atoms will change the sp$^2$ hybridization of carbon atoms to tetragonal sp$^3$ hybridization, modifying the surface geometry.[30, 31] Second, the impacts of heavy Ar ions, present in the plasma, could also modify the surface by inducing geometrical displacement of carbon atoms (rippling graphene surface) or creating vacancies and other defects. Figure 3(b) shows the topography image of the surfaces CVD graphene after the extended plasma treatment. The corrugation increases after the treatment and there are brighter regions in where the atomic resolution is lost or strongly distorted.

We have also found that these bright regions present a semiconducting behaviour while the rest of the surface remains conducting (see Figure 3(c)-(d)). Although the room temperature thermal drift makes it challenging to spatially resolve the electronic properties of the samples by STS with atomic accuracy, it is possible to analyze statistically thousands of tunneling spectroscopy measurements to infer changes on the electronic properties of the graphene surface (see Ref. [8] for more details).

### 3.2. Porphyrin physisorption

Iron porphyrins (FePP) are deposited onto the CVD graphene surface by drop casting. The porphyrins were used without any further purification from Aldrich. A solution 1.63 mM of FePP in uvasol grade chloroform was prepared. Further, a droplet of 10 µL of the porphyrin solution was dropcasted onto CVD-graphene surface and the solvent was left to evaporate for 3h in open air condition.

Figure 4a shows the STM topography of the CVD graphene surface after the deposition of FePP molecules. We observe six-pointed star shape protuberances with 0.26 nm in height and approximate dimension of 13 Å x 19 Å. These lateral dimensions are in agreement with the dimensions of a single porhyrin molecule obtained by first principle calculations (16 Å x 16 Å) (Hyperchem 6.0). The resolved six-star shape, however, does not perfectly match the expected geometrical structure of porphyrin molecules (four-star shape). The additional features in the observed star shape can be due to the splitting of the local density of the state (LDOS) of one phenyl ring. Another interesting finding is that FePP molecules are preferable physisorbed at the valleys of the rippled structure of graphene (Fig. 4a).

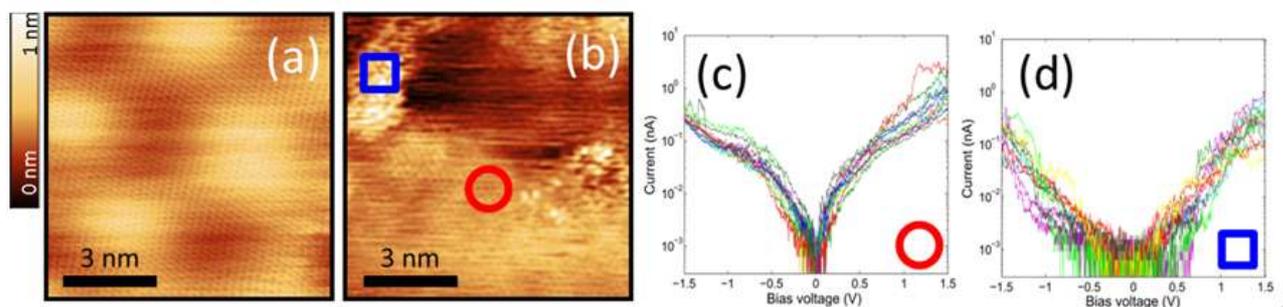

**Figure 3.** (a) and (b) are topography images, acquired in the constant current STM mode, of graphene grown by CVD on top of a nickel surface before and after hydrogenation. (c) Several current *vs.* voltage traces measured for a CVD graphene sample in regions with pristine atomic resolution like the one marked with the red circle in (b). (d) Same as (c) but measured in several bright regions, like the one marked with the blue square in (b), where the atomic resolution is distorted. Adapted from Ref. [8].
**Figura 3.** (a) y (b) son imágenes de topografía, adquiridas en modo de corriente constante, de grafeno crecido por CVD sobre níquel antes y después de hidrogenación. (c) Curvas corriente *vs.* voltaje medidas en varias regiones con resolución atómica como la marcada con el círculo rojo en (b). (d) Igual que (c) pero medidas en varias regiones brillantes, como la marcada con un cuadrado azul en (b), donde la resolución atómica está distorsionada. Adaptada de Ref. [8].

After the deposition of FePP onto the graphene surface, we measured the *I-V* traces directly on top of six porphyrin molecules (as the one indicated by the red cross in Figure 4a). The tunneling differential conductance shows a strong reduction in comparison to pristine graphene. In fact, the tunneling differential conductance traces present zero conductance (below the experimental resolution, 10 pS) for a wide range in bias, a characteristic of semiconducting behaviour while for pristine graphene the low bias differential conductance value is significantly larger than zero. Surprisingly, when this measurement is repeated on graphene several angstroms away from a single porphyrin molecule (at the topmost point of Moiré as indicated by the black cross in Fig 4a) the characteristic *I-V* traces (Fig. 4c) are different than that of pristine graphene (Fig. 4b). The tunnelling differential conductance at low bias is still strongly reduced. This indicates that the non-covalent modification of the graphene surface can be used to engineer its electronic properties.

## Conclusions

We have studied the local electronic properties of graphene using scanning probe microscopy techniques. We found that when graphene is deposited on a $SiO_2$ surface, the charged impurities present in the graphene-substrate interface produce strong inhomogeneities of the electronic properties of graphene. On the other hand, we shown how homogeneous graphene grown by CVD can be altered by chemical modification of its surface. We found that both the chemisorption of hydrogen and the physisorption of porphyrin molecules strongly depress the local conductance at low bias indicating the opening of a bandgap in graphene.

## Acknowledgements

A.C-G. acknowledges fellowship support from the Comunidad de Madrid (Spain) and the Universidad Autonoma de Madrid (Spain). M.W., and N.T. acknowledge financial support from the Ubbo Emmius program of the Groningen Graduate School of Science, the Zernike Institute for Advanced Materials and the Netherlands Organization for Scientific Research (NWO-CW) through a VENI grant. A. thanked financial support from the Foundation for Fundamental Research on Matter (FOM G-08). This work was supported by MINECO (Spain) through the programs MAT2008-01735, MAT2011-25046 and CONSOLIDER-INGENIO-2010 'Nanociencia Molecular' CSD-2007-00010, Comunidad de Madrid through program Nanobiomagnet S2009/MAT-1726.

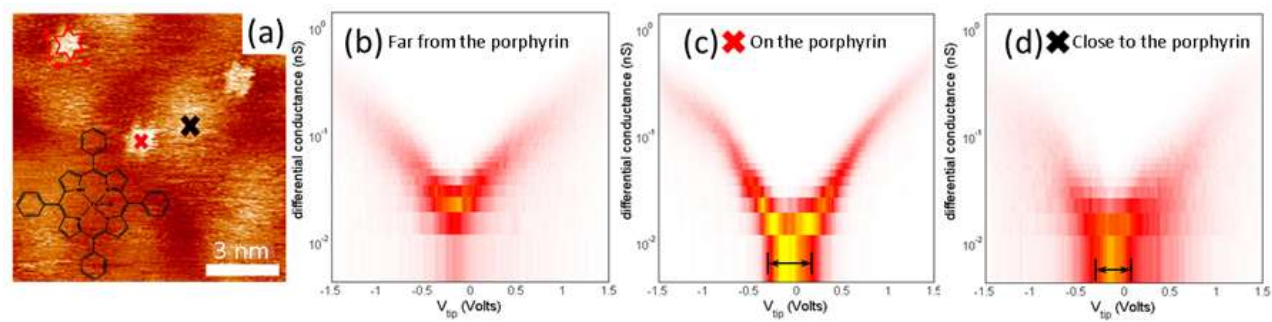

**Figure 4.** (a) 5,10,15,20-Tetraphenyl-21H,23H-porphine iron(III) chloride (abbreviated as FePP) molecules decorating the terraces of CVD graphene ($I_t$ = 0.1 nA; *V* = 0.5 V). FePP can be found at the valleys of the Moiré pattern of graphene. (b)-(d) Comparison of the differential conductance measured on the graphene sample after FePP deposition. The data is presented as 2D histograms, each one built from 1000 individual STS traces. (b) STS spectra measured on top of the Moiré pattern far from the molecules. (c) d*I*/d*V* traces acquired on top of the porphyrin molecule marked with a red cross in Fig. 4a. (d) STS measured on a hill of the Moiré close to a porphyrin molecule (black cross in Fig. 4a).

**Figura 4.** (a) 5,10,15,20-tetrafenil-21H,23H-porfirina de cloruro de hierro (III) (abreviado como FePP), decorando las terrazas de grafeno CVD ($I_t$ = 0.1 nA; *V* = 0.5 V) . Las FePP se pueden encontrar en los valles del patrón de Moiré del grafeno. (b) - (d) Comparación de la conductancia diferencial medida en la muestra después de la deposición de FePP en grafeno. Los datos se presentan como histogramas 2D, cada uno construido a partir de 1000 trazas STS individuales. (b) espectros STS medido en la parte superior del patrón de Moiré lejos de las moléculas. (c) Curvas d*I*/d*V* adquiridas en la parte superior de la molécula de porfirina marcada con una cruz roja en la Figura 4a. (d) STS medido en la parte superior del patrón de Moiré cercano a una molécula de porfirina (cruz de color negro en la figura. 4a).